\documentclass[aps,prd,twocolumn]{revtex4}

\usepackage[colorlinks=true, linkcolor=blue, citecolor=blue, urlcolor=blue]{hyperref}

\begin{document}

\title{A new search for the doubly charmed baryon $\Xi_{cc}^+$ at the LHC}

\author{Xing-Gang Wu}

\email{wuxg@cqu.edu.cn}
\affiliation{Department of Physics, Chongqing University, Chongqing 401331, People's Republic of China}

\date{\today}

\maketitle

The standard quark model is capable of predicting the existence of doubly heavy baryons. Similar to doubly heavy mesons, doubly heavy baryons may allow the QCD-inspired potential model, the nonrelativistic QCD (NRQCD) factorization theory~\cite{Bodwin:1994jh}, etc. to  work on them well, while serving as a fruitful ``laboratory" for testing these theories when a greatly enough collection of relevant data is available. Many attempts have been made to observe doubly heavy baryons, yet none of them succeeded due to the difficulties in producing such baryons either at the $e^+ e^-$ colliders or at the hadronic colliders. Noticeably, the running of the large hadronic collider (LHC) with a large centre-of-mass proton-proton collision energy and a much higher luminosity provides a good opportunity for experimentalists to realize the aforementioned observation.

In 2017, the doubly charmed baryon $\Xi^{++}_{cc}=ccu$ was firstly observed by the LHCb collaboration via the decay channel $\Xi^{++}_{cc}\to\Lambda_{c}^+ K^- \pi^+\pi^+$ with $\Lambda_{c}^+\to p K^-\pi^+$~\cite{Aaij:2017ueg}, using a data sample corresponding to an integrated luminosity of 1.7 fb$^{-1}$. This observation was confirmed by Ref.\cite{Aaij:2019zxa} and also by Ref.\cite{Aaij:2018gfl} via measuring another decay channel $\Xi^{++}_{cc}\to \Xi_{c}^{+}\pi^{+}$ with $\Xi_{c}^+\to p K^-\pi^+$, and its lifetime was measured too~\cite{Aaij:2018wzf}. Such observation demonstrates how really the LHC is a powerful discovery machine and confirms the correctness of the QCD-inspired quark model, stimulating a new world-wide trend for the theoretical studies on the doubly heavy flavored baryons.

Following the widely accepted hadronic production mechanisms for the doubly charmed baryon, firstly, a $(cc)$-diquark ``core" is perturbatively produced; and secondly, the $(cc)$-diquark core fragments into the final colorless $\Xi^{++}_{cc} (ccu)$ or $\Xi^{+}_{cc} (ccd)$ baryon by grabbing a $u$ or $d$ quark from the vacuum with the same probability~\cite{Sjostrand:2006za}. The production cross-sections of the doubly charmed baryons follow the probability of $\sigma(\Xi^{++}_{cc}) : \sigma(\Xi^{+}_{cc})= 1 :1 $. Thus the $\Xi^{+}_{cc}$ baryon is expected to be produced at the LHC as much as the $\Xi^{++}_{cc}$ baryon, and it is hoped that the $\Xi^{+}_{cc}$ baryon can also be observed at the LHC~\cite{Chang:2006eu}, though it is much harder to be tagged due to smaller branching ratio(s) of its observing channel(s) and its much shorter lifetime than that of the $\Xi^{++}_{cc}$ baryon~\cite{Yu:2017zst, Li:2017ndo}.

In 2002 and 2005, the SELEX Collaboration at the Tevatron collider claimed the observation of $\Xi^{+}_{cc}$ via the decay modes $\Xi_{cc}^+ \to \Lambda_c^+ K^- \pi^+$~\cite{Mattson:2002vu} and $\Xi_{cc}^+ \to p D^+ K^-$~\cite{Ocherashvili:2004hi}. However, such large $\Xi^{+}_{cc}$ production rates reported by the SELEX Collaboration have not been confirmed either by the FOCUS Collaboration~\cite{Ratti:2003ez} at the same collider or by the BABAR and BELLE Collaborations~\cite{Aubert:2006qw, Kato:2013ynr} at the $e^+ e^-$ colliders.

In 2013, the LHCb Collaboration performed the first search of the $\Xi_{cc}^+$ baryon at the LHC with their collected data sample at that time, corresponding to an integrated luminosity of 0.65 fb$^{-1}$ recorded at a centre-of-mass energy of 7 TeV; No significant signal was found. Recently, with the accumulated data sample at the centre-of-mass energies of 7, 8 and 13 TeV, corresponding to a total integrated luminosity of 9 fb$^{-1}$, the LHCb Collaboration performed a new search for the $\Xi_{cc}^+$ baryon~\cite{Aaij:2019jfq}. As an important step forward, the upper limit of ${\cal R}(\Lambda_c^+)$ was improved by an order of magnitude than the previous search, however still no significant signal has been observed. Here ${\cal R}(\Lambda_c^+)$ is the ratio of the production cross-sections between the $\Xi_{cc}^{+}$ and $\Lambda_c^+$ baryons times the branching fraction of the $\Xi_{cc}^{+} \to \Lambda_c^+ K^- \pi^+$ decay. The $\Xi^+_{cc}$ selection efficiency and hence the ratio ${\cal R}(\Lambda_c^+)$ depend strongly on the lifetime. Since the $\Xi^{+}_{cc}$ baryon has a much smaller lifetime than that of the $\Xi^{++}_{cc}$ baryon, it is much harder to be tagged from the background by using the vertex detector (at the LHCb). Then more $\Xi^{+}_{cc}$ events than $\Xi^{++}_{cc}$ events are lost, and the characteristic decay channels of $\Xi^{+}_{cc}$ have much smaller branching ratios than those of $\Xi^{++}_{cc}$. These facts may explain why $\Xi^{+}_{cc}$ is much harder to be observed than $\Xi^{++}_{cc}$ at the LHC. We can expect that future LHCb searches with further improved record conditions, additional $\Xi^{+}_{cc}$ decay modes can be used; if larger data samples are collected, it can significantly increase the $\Xi^{+}_{cc}$ signal sensitivity.

One may observe that the measured ratio ${\cal R}(\Lambda_c^+)$~\cite{Aaij:2019jfq} is significantly below the value reported by the SELEX Collaboration. Because the intrinsic or extrinsic charm components in a proton could have significant contributions to the productions of the doubly heavy baryons in the small $p_t$ region~\cite{Chang:2006xp}, we still need more data to confirm whether the SELEX result is correct or not, or maybe the SELEX and LHCb measurements are consistent with each other~\cite{Brodsky:2017ntu}. The future experiments at the RHIC~\cite{Chen:2018koh}, or the LHeC~\cite{Huan-Yu:2017emk}, or the fixed-experiment at the LHC (After@LHC)~\cite{Chen:2014hqa}, or the ILC~\cite{Chen:2014frw}, may have some help. Moreover, careful theoretical studies on the doubly heavy baryon decays are also important for us to clarify the issue.

\end{document}